\documentclass[
reprint,
amsmath,amssymb,
aps,
]{revtex4-1}

\usepackage{float}
\usepackage{graphicx}
\usepackage{dcolumn}
\usepackage{bm}
\usepackage{lineno}
\usepackage{hyperref}

\begin{document}

\title{An apparatus architecture for femtosecond transmission electron microscopy}
	
\author{C. W. Barlow Myers}
\author{N. J. Pine}
\author{W. A. Bryan}
\email{w.a.bryan@swansea.ac.uk}
\affiliation{Department of Physics, College of Science, Swansea University, Singleton Park, Swansea SA2 8PP, UK}
	
\date{\today}
	
\begin{abstract}
The motion of electrons in or near solids, liquids and gases can be tracked by forcing their ejection with attosecond x-ray pulses, derived from femtosecond lasers. The momentum of these emitted electrons carries the imprint of the electronic state. Aberration corrected transmission electron microscopes have observed individual atoms, and have sufficient energy sensitivity to quantify atom bonding and electronic configurations. Recent developments in ultrafast electron microscopy and diffraction  indicate that spatial and temporal information can be collected simultaneously. In the present work, we push the capability of femtosecond transmission electron microscopy (fs-TEM) towards that of the state of the art in ultrafast lasers and electron microscopes. This is anticipated to facilitate unprecedented elucidation of physical, chemical and biological structural dynamics on electronic time and length scales. The fs-TEM numerically studied employs a nanotip source, electrostatic acceleration to 70 keV, magnetic lens beam transport and focusing, a condenser-objective around the sample and a terahertz temporal compressor, including space charge effects during propagation. With electron emission equivalent to a 20 fs laser pulse, we find a spatial resolution below 10 nm and a temporal resolution of below 10 fs will be feasible for pulses comprised of on average 20 electrons. The influence of a transverse electric field at the sample is modelled, indicating that a field of 1 V/$\mu$m can be resolved.

\end{abstract}
	
\maketitle
\section{Introduction}

Direct observation of the time evolution of ultrafast processes is challenging due to the characteristic small length and time scales over which these dynamics occur. Nonetheless, it is achievable providing one has a probe of sufficient brightness, short wavelength and temporal duration: electron matter waves meet these requirements and fit on a table-top. The coupling of the spatial resolution achievable in transmission electron microscopy (TEM) \cite{Palatinus2017,Nellist2004,Batson2002} with the temporal resolution of femtosecond (fs) laser technology was pioneered by Zewail et al with picosecond (ps) resolution \cite{Zewail2000,Williamson1997,ihee1} and later extended into the fs regime by Miller et al \citep{Siwick2003}. This provides a route to achieving the ultimate goal of ultrafast imaging: the `molecular movie' \cite{Sciaini2011}. Observations of other processes facilitated by ultrafast electron imaging include phonon dynamics \cite{siwickgraphene,Cremons2016, Park2005, Gao2012}, laser induced ablation \cite{Carbone2011}, probing evanescent fields near nanostructures \cite{Barwick2009, Feist2015,Piazza2015,ryabov2016,Quinonez2013a,Yurtsever2012}, chemical reactions \cite{Ihee2001,Srinivasan2005, gao}, laser induced magnetism \cite{4DEM}, phase transitions \cite{Baum2007,Haupt2016,Ruan2007,Siwick2003,Wall2012} and transient molecular alignment \cite{Reckenthaeler2009}. 

A number of numerical and experimental studies have produced nanometre and femtosecond resolutions, but not simultaneously. Compact electron imaging systems utilising passive dispersion control have reached temporal resolutions of the order of 100fs \cite{Gerbig2015,Badali2016,Feist2017,Quinonez2013a} and can reach resolutions down to 20 fs (FWHM) when optimised \cite{Hoffrogge2014a}. Simulations of active compression of single electron pulses to sub-10 fs have been conducted \cite{Veisz2007,Fill2006} albeit with no discussion of spatial or field resolution. Active compression can produce multi-electron pulses of the order of 10 fs, however with temporal resolution limited to tens of femtoseconds due to timing instabilities \cite{Maxson2017}. Lensless point projection microscopy \cite{fssingletofew} paired with plasmon driven electron emission from a structured nanoscale metal tip  (NSMT) \cite{Muller2016} could permit comparable temporal and spatial resolutions due to potentially nanometer source-sample distances, limited by the time-bandwidth effects of the driving laser. 

In this paper we present the numerical modelling of a novel femtosecond transmission electron microscope (fs-TEM), consisting of a NSMT photoelectron source, electron optics, and detection. Temporal dispersion is compensated both passively and actively. Studies of each beamline component are presented in the context of preservation of pulse duration from photoemission to sample. A magnetic condenser objective lens gives full control of the beam diameter at the sample and ensures that the angular velocity of the pulse is eliminated when the pulse traverses the plane of the sample, with magnifications in excess of $10^4$ possible.  Comparison of the distribution at the sample and detector planes infers minimal spatial distortion. The minimum resolvable electric field is explored via time dependant electric fields superimposed on a nanometrically structured sample.

\section{Results}
\subsection{Photocathode Emission and Geometry }

As typical with NSMT sources, our fs-TEM consists of polycrystalline tungsten wire (125 $\mu$m radius) which tapers to an apex with a 50 nm radius of curvature and an opening angle of 10 degrees, as shown in figure 1 (a). Following the work of Hoffrogge et al \cite{Hoffrogge2014a}, Paarmaan et al \cite{Paarmann2012} and Bormann et al \cite{Bormann2015}, we incorporate an electrostatic suppressor, a 50 $\mu$m thick annulus with an inner diameter of 150 $\mu$m. The NSMT protrudes 100 $\mu$m from the suppressor surface, sufficient space for laser excitation. Grating coupled laser driven electron emission from a NSMT has also recently been demonstrated, relaxing the requirement for laser access.

\begin{figure*}
	\includegraphics[width=\textwidth]{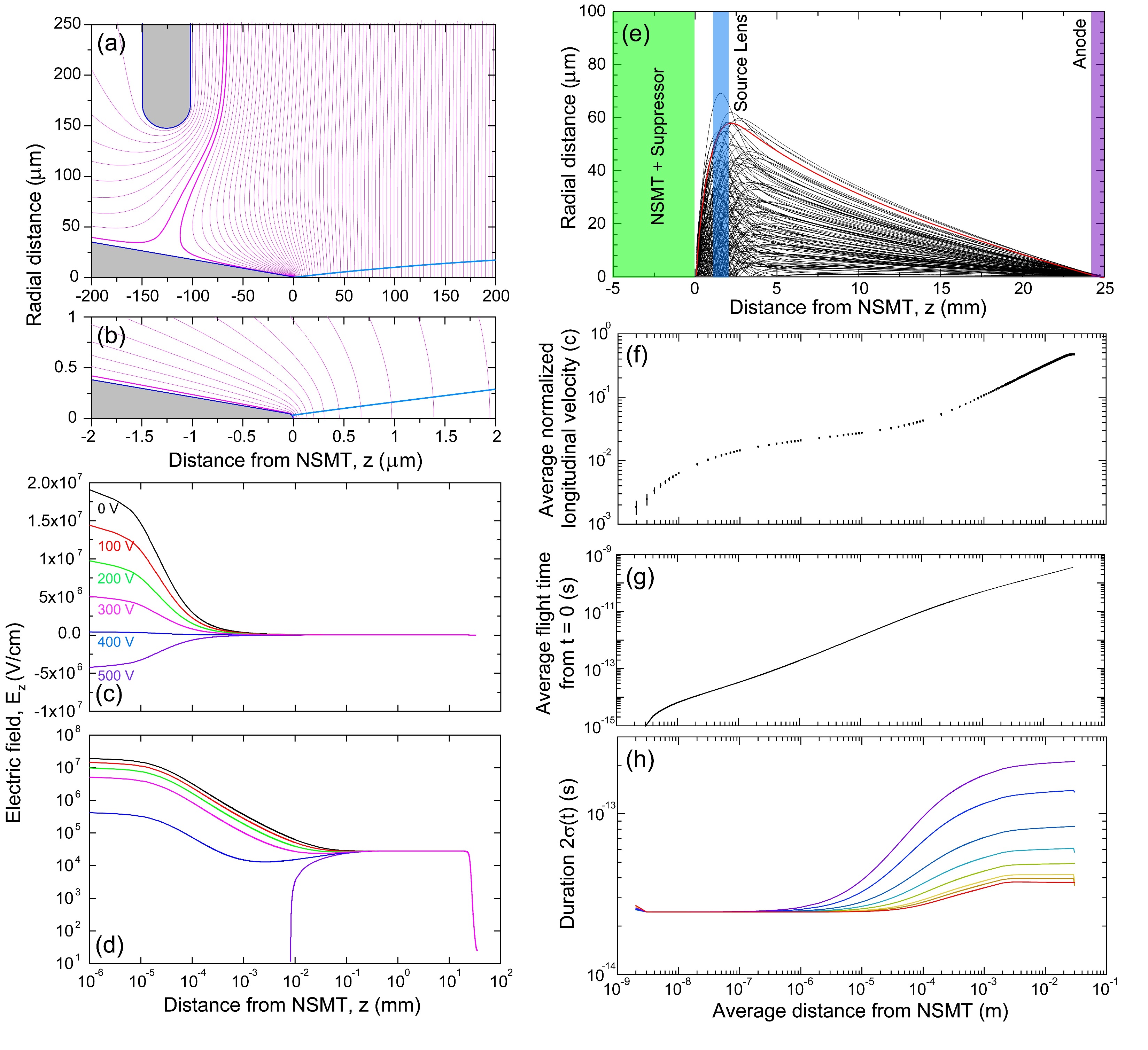}
\end{figure*}
	
\begin{figure*}
	\caption{(a) Contour plot of the static (DC) electric field in the vicinity of the NSMT and the suppressor parallel (z-axis) and perpendicular (r-axis) to the axis of cylindrical symmetry. The saddle induced by the suppressor indicated by thick magenta lines, and the 2$\sigma$ outer trajectory of the electron pulse is shown in blue. (b) Expanded view of the apex of the NSMT. The thick magenta line again indicates the saddle point which attaches to the NSMT just behind the apex. In (a) and (b), $V_{sup}$ = 70.1 kV and $V_{tip}$ = 70 kV, and the contour lines are separated by 10 V. (c) DC electric field along the z-axis (E$_z$) induced between combination of the NSMT (at potential $V_{tip}$ = -70 kV) and suppressor (at $V_{sup}$) and the grounded anode at 2.5 cm as $\Delta V = V_{tip}-V_{sup}$ is varied between 0 and 500 V, shown on a linear scale. (d) as (c) with $E_z(z)$ on a logarithmic scale. (e) Individual simulation macroparticle trajectories from the NSMT and Suppressor, through the maximum field of the source magnetic lens (blue) and exiting through an aperture (radius 1 mm with 0.5 mm fillet) in the grounded anode (purple). The red line indicates 2$\sigma$ of the position coordinates of all macroparticles. (f) Longitudinal velocity of the electron pulse averaged over all GPT macroparticles as a function of average z position on a logarithmic scale. For a potential difference of 70 kV, a final velocity of 0.48c is anticipated. (g) Average time of flight from creation at time t = 0 as a function of average z position. (h) The 2$\sigma$(t) electron pulse duration as a function of average z position. The colour assignment from the lowest 2$\sigma$(t) is zero charge, then respectively 1e, 2e, 5e, 10e, 20e, 50e and 100 electrons per pulse. This colour coding is reused throughout.}
	\label{fig:fig1}
\end{figure*}

We apply a 25 $\mu$m fillet to the inner circumference of the suppressor to prevent high field gradients causing field emission. The tip-anode distance is fixed at 2.5 cm throughout, such that the surface fields on all electrodes are below 5 MV/m. The nanotip potential ($V_{tip}$) and suppressor potential ($V_{sup}$) define the surface fields at the emission site. With the suppressor set at a potential comparable to the tip ($\Delta V = V_{sup} - V_{tip} \approx 0$) there is a pronounced bunching of the equipotentials around the apex, shown in figures 1(a) \& (b), indicating a strong DC gradient. The dominant gradient lies along the z-axis towards the anode and the on-axis field E$_z$(z) is explored as a function of $\Delta V$, shown in figure 1(c) \& (d). Increasing $\Delta V$ moves the saddle point (highlighted in the equipotential lines of figure 1(a) in dark magenta), towards the apex of the tip and ultimately into the z $\geq$ 0 region. This causes field reversal and an extinction of the photo-electron current, present in figure 1(c) \& (d) for $\Delta V$ $>$ 400 V. 

Changing $\Delta V$ modifies the electrostatic landscape around the NSMT, facilitating adjustment of the spatio-temporal distribution of the electron pulses \cite{Paarmann2012}. The reduction of the DC field decreases the longitudinal momentum acquired during acceleration (see figure 1(c) \& (d)), increasing the time required to reach the sample and increasing propagation induced temporal broadening. The reduction of the DC field gradient reduces the transverse momentum of electrons and hence reduces the path length differences for electrons emitted across differing angles with respect to the z axis. These two processes cannot be varied independently. Typically, the DC electric field would be set as high as possible while suppressing field emission. With $V_{tip} = -70$ kV, $V_{sup}$ would be set such that the surface field of the NSMT does not exceed 1.5 GV/m \cite{Yanasigawa2010}, hence optical field enhancement would only permit electron tunnelling during the presence of the driving laser pulse. In practice, this would depend on the microscopic shape of the NSMT, and would be the first parameter varied when identifying operating conditions.

The region between the NSMT and the anode is shown in figure 1(e). The inclusion of a magnetic lens allows the divergence of the trajectories from the NSMT to be controlled independently of $\Delta V$. This lens focusses the electron pulse at the aperture of the anode, ensuring distortions induced by fringing fields are minimal, also reducing spherical aberrations. Geometric temporal expansion of the pulse is also minimized by reducing path length differences across electron trajectories.

Electron pulse rotation about the propagation direction (z-axis) induced by a magnetic lens depends on the longitudinal velocity. In the acceleration region, the subsequent focusing is coupled to the DC field gradient, and the best performance is varied is found when the electron bunch has reached a significant proportion of its final velocity. This is shown through a comparison between figures 1(e) and (f). For a final beam energy of 70 keV, a maximum radius of 2$\sigma$(r) = 55 $\mu$m is achieved, where $\sigma$ is the standard deviation of the parameter in parenthesis, and is used throughout. The optimal balance between minimized 2$\sigma$(r) and a small temporal geometric stretch is found when a strong magnetic field is applied in the first few millimetres of electron flight. The convergence angle of 2.5 mrad observed in figure 1(e) is at least an order of magnitude smaller than the divergence angle when no source magnetic field is present. 

Comparing figures 1(c) and 1(f) at small z position (z $<$ 10 nm), the DC electric field is relatively constant over the first ten nanometres, causing near-constant acceleration. Between 10 nm $<$ z $<$ 100 $\mu$m, the electric field transitions between being governed by the NSMT to the parallel plate case. This is shown in figure 1(d), by the tendency for the different V cases to converge to a constant value at z $>$ 100 $\mu$m. Examination of the flatness of the contours in figure 1(a) around z $>$ 100 $\mu$m, the linearity of the electric field E$_z$ in figure 1(d) and the linear increase of velocity with z position in figure 1(f) indicates the completion of the transition between NSMT-influenced acceleration and parallel plate acceleration. Upon exit of the anode the electrons reach their final velocity of $v_z \approx$ 0.48c ($\lambda_{e} \approx $ 4.5 pm), as indicated in figure 1(f).

The total flight time of the electron pulse from the NSMT to the anode is shown in figure 1(g). The significant DC field gradient at the apex of the tip means that the electron pulse accelerates dramatically in the first 10 nm, which is a key consideration when trying to mitigate the influence of space charge. The transnational behaviour detailed above is also apparent from the flight time. In the case of holding the NSMT at 70 kV, the total acceleration time is around 400 ps. 

The influence of space charge on the 2$\sigma$(t) pulse duration is further shown in figure 1(h). The duration of electron release is 2$\sigma$(t) = 25 fs, which including geometric effects, is comparable to that induced by a 20 fs FWHM laser pulse. A charge of one to five electrons increases 2$\sigma$(t) from 25 to 40 fs. A charge of tens of electrons results in a major temporal stretch, exceeding 100 fs with fifty electrons per pulse. With expected temporal compression ratios of the order of 10, exceeding tens of electrons per pulse by a large margin will result in a pulse that will be difficult to compress to low tens of femtoseconds duration.

\subsection{fs-TEM electron optics configuration}

\begin{figure*}
	\includegraphics[scale=0.35]{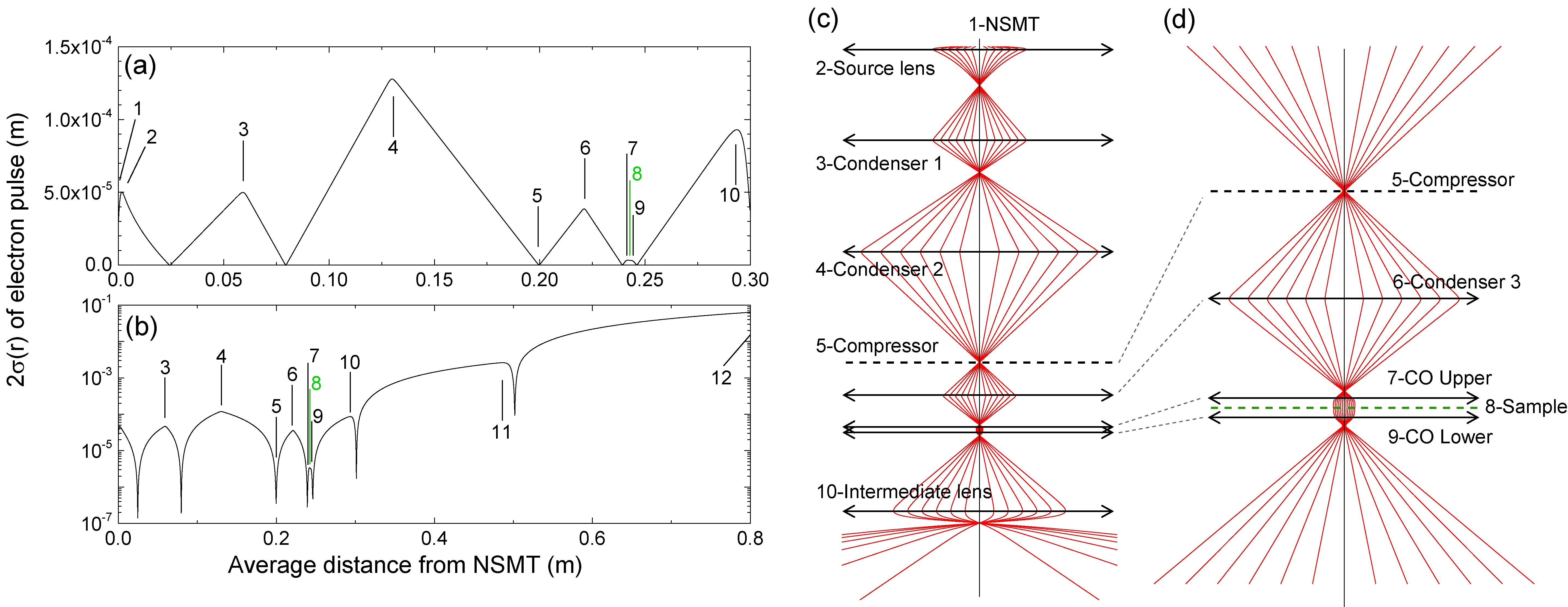}
	\caption{Configuration of the modelled fs-TEM instrument with elements identified: (1) NSMT electron source, (2) source lens, (3) condenser lens 1 (C1), (4) condenser lens 2 (C2), (5) temporal compressor, (6) condenser lens 3 (C3), (7) upper part of condenser-objective magnetic lens (CO1), (8) sample plane, (9) lower part of condenser-objective (CO2), (10) intermediate lens, (11) projector lens and (12) electron detection plane. (a) 2$\sigma$(r) of the electron pulse as it propagates through the fs-TEM, where r is radial position in the x,y plane. (b) as (a) but over the full z-axis with 2$\sigma$(r) on a logarithmic scale. In (a) and (b) there is no electron pulse charge. As the pulse charge is increased, C1-3 and CO lens strengths are adjusted. (c) Schematic of the electron pulse trajectory from the source to intermediate lens, where the back dashed line indicates the position of the temporal compressor at the focus of condenser 2. (d) as (c) but highlighting the condenser-objective and sample plane (green dashed line).}
	\label{fig:fig2}
\end{figure*}

Defining design criteria for a high performance fs-TEM source is relatively straightforward, being low or controllable divergence, short pulse duration, minimal space-charge distortion, high coherence and excellent beam quality. Defining similar criteria for the electron optics between the source and sample requires a series of often contradictory conditions to be met simultaneously. Our design criteria for the fs-TEM are defined in Methods. Following a series of iterations, we present a fs-TEM design in Fig \ref{fig:fig2}, incorporating the source discussed above and in accordance with these five criteria. 

Summing the radius of the electron pulse Airy disk and the spherical aberration in quadrature gives an indication of spatial resolution \cite{williamsandcarter}. For an electron wavelength of 4.5 pm and assuming a spherical aberration coefficient $C_{S}$ = 3 mm, an optimal convergence angle of 4.5 mrad is found, which corresponds to an optimal resolution of 0.7 nm. However, an electron pulse from a NSMT has a degree of inherent curvature which increases the apparent $C_{S} \geq$ 10 mm, increasing the optimal resolution to greater than 1 nm. There is a possibility that this could be improved with multipole aberration correction, however such devices add to the flight length of the instrument, hence gains in spatial resolution would compromise temporal resolution.

Temporal control is implemented both passively and actively: passive control via considerations of the space-charge influence of each component and minimization of the total source to sample distance, and active control via a laser-driven resonator. The resonator acts by modifying the position-velocity distribution such that a temporal focus occurs at the sample plane. The beam is spatially focussed at the resonator, indicated in figure 2, to mitigate the effects of field imperfections. 

For maximum instrument flexibility, parallel and convergent beams of varying diameter are needed at the sample which is impossible to achieve with a standard magnetic lens. Condenser-Objective (CO) lenses, established in TEM for four decades, see a converging beam passing through two (upper and lower) strong lenses acting to collimate the beam at a radius 2$\sigma$(r) on the micron scale. Changing the C3 lens strength allows non-parallel illumination such that spot size and convergence angle are independent.

Post-sample optics consist of intermediate and projector lenses, providing magnification and delivering the beam with appropriate dimensions to the electron detector. As temporal information is encoded by a pump-probe measurement, following delivery of a short electron pulse to the sample, the evolution of the electron image as a function of delay requires only that the post-sample optics preserve the image, not the temporal characteristics. Furthermore, the action of the compressor will cause a strong temporal divergence, hence the charge density drops. This results in the post-sample electron transport optics acting in a manner akin to a TEM.

\subsection{Propagating ultrafast electron pulses through magnetic lenses}

As shown in the fs-TEM schematic in figure 2, the electron pulses are sent through four spatial foci before interaction with the sample, therefore qualification of any resulting spatio-temporal distortion is necessary \cite{Weninger2012}. As the pulse passes through a spatial focus, the beam dimensions and hence severity of temporal expansion depend on the incident pulse duration, number of electrons and convergence angle. We demonstrate this in figure 3, which shows the influence of convergence angle and electron pulse charge on temporal duration. Rather than consider the full fs-TEM as shown in figure 2, in the case of figure 3, we start a collimated electron pulse at z = 0 with an average energy of 70 keV and\cite{Schneider2014}  a bandwidth of 0.5 eV. The 2$\sigma$(r) radius of the electron pulse is 50 $\mu$m, comparable to that found in figure 2.

\begin{figure}
	\centering
	\includegraphics[scale=0.4]{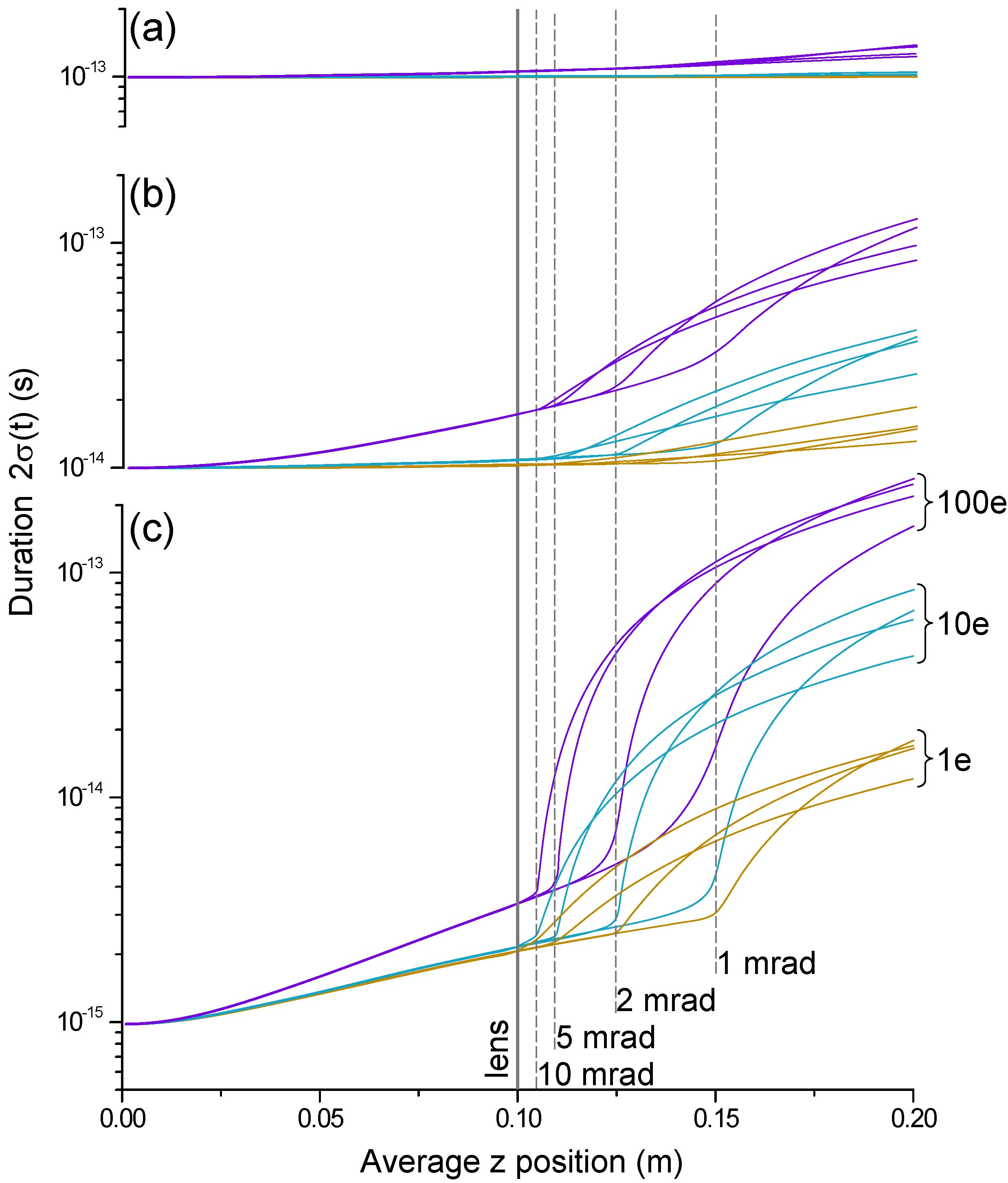}
	\caption{Temporal action of a range of magnetic lens strengths on a 70 keV electron pulse of varying initial duration and charge. Note this electron pulse is generated at z = 0 at 70 keV unlike all other figures. The magnetic lens is at z = 0.1 m (solid line) and forms foci at z = 0.15 m (convergence angle of 1 mrad), z = 0.125 m (2 mrad), z = 0.11 m (5 mrad) and z = 0.105 m (10 mrad), indicated by dashed lines. (a) Initial duration of 2$\sigma$(t) = 100 fs. (b) Initial duration of 2$\sigma$(t) = 10 fs. (c) Initial duration of 2$\sigma$(t) = 1 fs. The 2$\sigma$(t) scale is comparable between panels. }
	\label{fig:fig3}
\end{figure}

Electrons pulses are created with 1, 10 and 100 electrons per pulse, for durations of 100, 10 and 1 fs, shown in figure 3(a - c) respectively. These 70 keV pulses are initially collimated and a solenoid lens at z = 0.1 m focuses with convergence angles varying from 1 to 10 mrad. Some temporal stretching is observed between z = 0 and 0.1 m, and is most dramatic for the shortest pulse and highest charge. 

In the case of the longest electron pulses, as apparent in figure 3(a), 100 fs electron pulses are distorted a relatively moderate amount. The electron pulse is observed to increase in duration, with the most obvious influence on the pulse containing the highest charge. As seen in figure 3(b), a 10 fs pulse with charge between 1 and 100e increases in duration by no more than an order of magnitude over the z-axis range we consider. Nonetheless, it is clear that propagating an electron pulse containing 100 electrons will be very challenging if we want to react a duration of below ten femtoseconds. 

For the 1 fs pulse the extreme longitudinal confinement coupled with radial confinement (spatial focus) causes a rapid explosion in time. As the convergence angle is decreased from 10 mrad to 1 mrad, the duration is seen to increase more rapidly with increasing z. Furthermore, for low convergence foci and 10 and 100e pulses, the duration is observed to be distorted over a far larger distance as compared to the high convergence foci. 

With an eye to design criteria (2 \& 3) (see Methods), we find that an electron pulse containing around 10 electrons, with an initial duration of 10 fs provides a good balance between minimisation spatial focus induced  temporal distortion and charge per pulse. Indeed, such a pulse repeatedly passing through foci would undergo most temporal distortion after the first focus, followed by successively reducing distortions as the duration further increases. As seen later, a pulse with a duration of high tens of femtoseconds is compressible to the extent that (3) will still be met.

While challenging, producing magnetic lenses is essentially a solved engineering problem. In condenser lenses C1 to C3, intermediate and projector lenses, the maximum field utilised is significantly less than 1 Tesla, far from saturating any components of the magnetic circuit. The condenser-objective requires a higher field, reaching 1.5 T in the present work, however it should be considered that this field is generated by a pair of single turn solenoids without a pole piece. It is therefore anticipated that lenses of a few thousand amp-turns and pole pieces constructed from a modern high performance soft metal alloy will allow such magnetic fields to be achieved without saturation.

\subsection{Temporal Compression}

To deliver 10 fs pulses over the flight distance necessary to include magnetic lenses, active compression is required. In \cite{Kealhofer2016} the compression and characterisation of ultrafast electron pulses using an aluminium resonator excited by 0.3 THz pulses was demonstrated. Optical field enhancement around the resonator yields electric fields of 10$^6$ V/m, compressing picosecond electron pulses by a factor of 12 over a distance of tens of centimetres. Timing issues are inherently minimized as the laser which drives the photoemission is also the pump of the optical rectification process \cite{Schneider2014} generating the THz pulses, resulting in a timing stability of better than 4 fs. 

\begin{figure*}
	\centering 
	\includegraphics[scale=0.85]{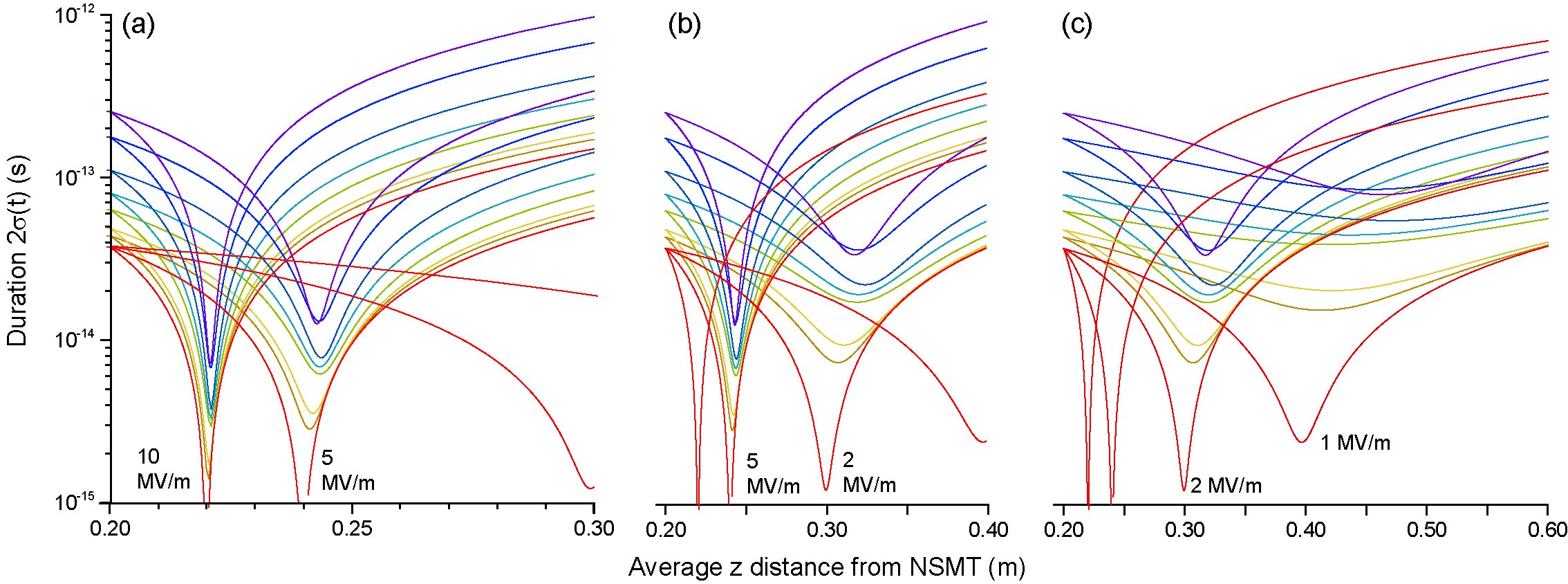}
	\caption{Active temporal compression mimicking the action of a terahertz resonator with varying peak field strength and electron pulse charge. The electron pulse is generated at the NSMT as shown in figure 1, and propagated to the compressor through condenser lenses 1 and 2, identified in figure 2. In all cases, the compressor is at z = 0.2 m from the NSMT. (a) Temporal foci for a peak field of 10 and 5 MV/m with an electron pulse charge varying from 0, 1e, 2e, 5e, 10e, 20e, 50e and 100e from lowest to highest 2$\sigma$(t). The red lines to the bottom right are a visual link to parts (b) and (c). (b) as (a) for 5 and 2 MV/m and (c) as (a) for 2 and 1 MV/m. For all fields and electron pulse charges, the strength of lenses C1 and C2 and the phase of the terahertz field are optimised for minimum duration.}
	\label{fig:fig4}
\end{figure*}

Rather than numerically solve Maxwell's equations for a THz resonator, we include the influence of the compressor by a time dependent momentum transfer to the electron pulse. This is discussed in the Methods. The action of the compressor is shown in figure 4 as the number of electrons per pulse and the peak electric field are varied. The changing electron density shifts the spatial foci of lenses C1 and C2, so all pre-compressor optics are optimised for each pulse charge. As with earlier figures, the duration of the electron pulse is quantified as 2$\sigma$(t). 

At a peak electric field of 10 MV/m as shown in figure 4(a), pulses of up to 100e can be compressed to a 2$\sigma$(t) duration less than 10 fs. A peak field of 5 MV/m is less experimentally demanding, and as is apparent from figure 4(a), electron pulses with a charge up to 20e can be compressed to a 2$\sigma$(t) less than 10 fs. As shown in figure 4(b), a decrease in the peak field to 2 MV/m causes a further increase in duration, whereby 1e and 2e pulses exhibit a duration of around 10 fs. This trend continues in figure 4(c), whereby the 100e pulses approach a minimum of only 100 fs. 

The shifting z-position of the minimum duration with increasing pulse charge is a consequence of the spatio-temporal variation of the charge density, influencing convergence and vice versa, with the outcome a varying compressivity. The peak electric field also directly defines the position of the temporal focus: increasing the peak field moves the temporal focus close to the compressor. This is analogous to the spatial focusing of the electron beam with magnetic lenses discussed earlier. We select a field strength of 5 MV/m, which is an achievable compromise between compressor performance and mechanical feasibility.

Fields in excess of 10 MV/m could compress pulses to shorter durations within the restriction placed by the breakdown limit of resonator material. Simulations of alternative resonator geometries have shown field enhancement factors which would facilitate such field strengths \cite{Bagiante2015}. However the electron optics required to collimate or focus the diverging beam onto the sample constricts the distance between the resonator and sample, necessitating a heavily modified fs-TEM geometry. Consideration must also be given to the laser optics required to define illumination of the sample, which is anticipated to be collinear with the electron pulse axis. 

\subsection{Condenser-objective lens implementation in fs-TEM} 
Manipulating the electron pulse spot size and convergence angle at the sample facilitates real or reciprocal imaging modes, and controls magnification. Optimal illumination minimises imaging time, crucial when the stability of the apparatus or sample places temporal limitations upon image acquisition. When utilising single or few electron pulses to remove or alleviate space charge induced pulse broadening, the matching of beam dimensions to sample dimensions is essential to ensure practical imaging times.

\begin{figure*}
	\includegraphics[width=\textwidth]{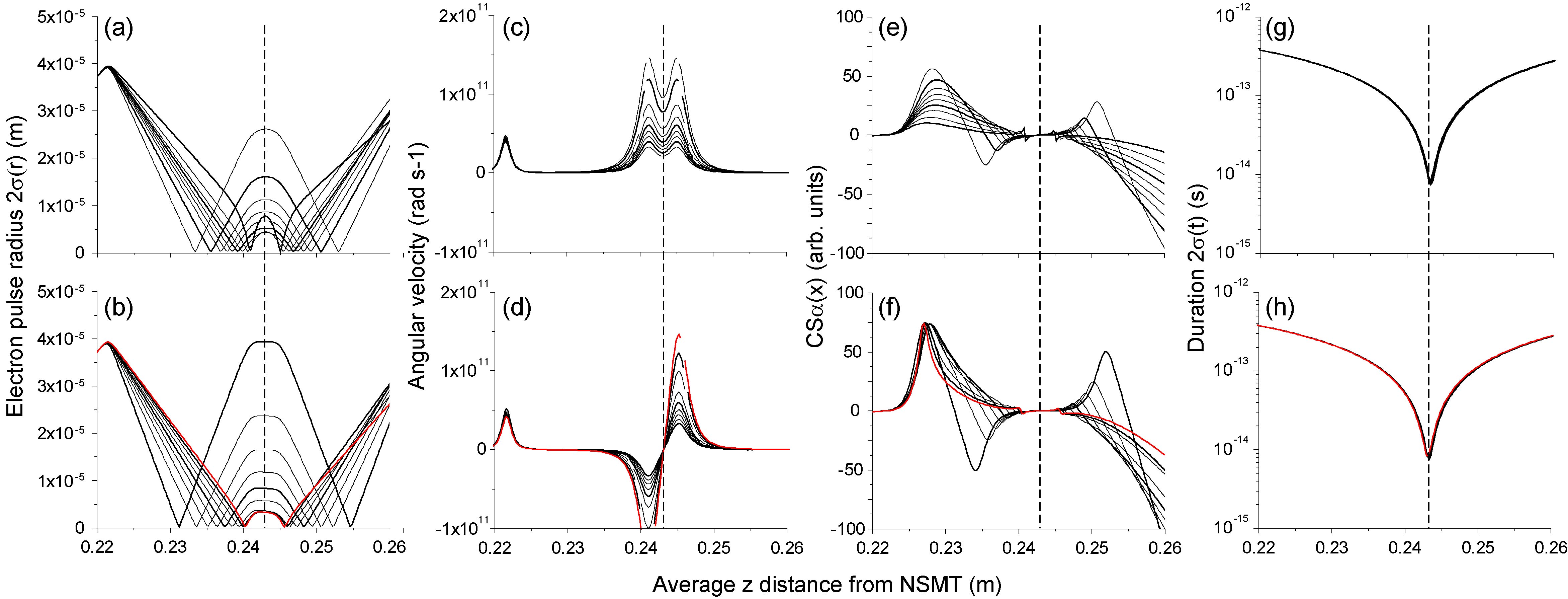}
	\caption{Spatial and temporal performance of the fs-TEM in the vicinity of the sample (at z = 0.243 m, indicated by vertical dashed line), controlled by the strength of condenser lens 3 and the upper and lower parts of the condenser-objective lens as identified in figure 2. The top panels are parallel magnetic lens excitation whereby C3, and both the upper and lower elements of the CO all have identical polarities, and the bottom panels are for anti-parallel excitation whereby the polarity alternates between C3 and the upper CO, and between the upper and lower CO elements. (a, b) 2$\sigma$(r) beam radius, (c, d) average angular velocity of the electron pulse, (e, f) Cournat-Snyder (or Twiss) alpha parameter and (g, h) 2$\sigma$(t) temporal duration. The bold lines are referenced to the filled points in figure 9 (shown in Methods), and the red line indicates the lowest 2$\sigma$(r) achieved.}
	\label{fig:fig5}
\end{figure*}

Placing the sample at the focus of a condenser lens restricts the fs-TEM system, as it is impossible to alternate between spatially focused or collimated illumination without major modification to the lens pole pieces. Furthermore, a highly divergent beam is favourable on exit of the sample plane to achieve high magnification, which is difficult to achieve in a traditional condenser lens.

In TEM, a condenser-objective (CO) lens combines the transport of the incoming electron pulse to the sample while also maximizing the post-sample divergence. A typical CO sees the sample placed equidistant between two strong magnetic windings, with B-fields approaching saturation in the soft metal circuits. These coils are excited by the same current, hence the symmetry of the lens collimates the beam at the sample. CO lenses are often coupled with an additional small condenser, manipulating the incoming convergence angle before the sample, facilitating convergent or parallel illumination without changing the CO strength. 

To demonstrate the applicability of the CO lens to fs-TEM, we include a symmetric twin-field system with the sample equidistant between the condenser and the objective single turn solenoids as shown in figure 2. Again, we do not include pole pieces in this design, rather take care so as maintain field strengths and gradients that can be managed with water-cooled lenses. We make use of the equivalent lens concept, whereby two solenoids of radius r separated by 2r is a good approximation of the field geometry in a CO lens.

The combined influence of C3 and the CO is tested under two operating conditions, parallel and antiparallel excitation polarities. Referring to figure 2, C3 and the upper and lower parts of the CO have the same polarity in the former, and C3 and the CO lower are excited with opposite polarities to the CO upper in the latter. The diverging electron pulse passes C3 at z = 0.2215 m, converging towards the CO upper lens at z = 0.238 m. The combined action of C3 and the CO upper collimates the electron pulse. The election pulse then passes a second focus after the sample as controlled by the CO lower (z = 0.248 m), either by continuing to rotate about the z-axis or reversing direction. 

Each C3 and CO combination in figure 5 is identified by setting the CO current (see figure 9 in Methods), then changing C3 until the electron pulse is collimated at the sample plane. All results are for a twenty electron pulse with an initial duration of 20 fs, space-charge is calculated without approximations over the 250 macroparticles used, and all lenses and the compressor are optimised as C3 and CO are adjusted. It is found that a focused electron pulse can also be formed at the sample plane, however we limit our discussion to parallel illumination. Systematically varying the C3 and CO currents changes the size of the collimated electron pulse at the sample plane. 

A range of collimated parallel and anti-parallel electron pulses are shown in figures 5(a) and (b) respectively. The degree of collimation around the sample is higher in the anti-parallel case. Figures 5(c) and (d) show the average angular velocity (proportional to the magnetic field) of the electron pulse in parallel and anti-parallel configurations. Parallel excitation is straightforward to achieve as the gradients required are less than in the anti-parallel case. Nonetheless, anti-parallel excitation has the significant advantage of forming a magnetic field zero at the sample plane, an advantage when observing laser induced magnetism. The gradient between the minimum and maximum angular velocities (hence magnetic field gradients) in figure 5(d) is approximately linear, causing the flattening of 2$\sigma$(r) around the sample shown in figure 5(b).

The collimation characteristics of the electron pulse in parallel and anti-parallel cases are shown in figures 5(e) and (f) respectively. The Courant-Snyder, or Twiss parameters ($\alpha$, $\beta$ and $\gamma$) describe the root-mean square (RMS) of the emittance of the electron pulse. The $\alpha$ parameter (here CS$\alpha$) describes the correlation of the RMS particle position and direction. When CS$\alpha$ = 0, the electron pulse is either collimated or at a spatial focus. In figures 5(e) and (f), CS$\alpha$ crosses zero at the sample plane. 

The action of the compressor is shown in figure 5(g) and (h), whereby the THz field strength and phase are set to achieve a temporal focus in the sample plane. The combined action of the CO and compressor is therefore a collimated electron pulse with 2$\sigma$(r) = 3 $\mu$m, and in the case of these 20e pulses, a 2$\sigma$(t) duration of 8.2 fs. Importantly, the influence of the CO and compressor are essentially independent, as apparent from figure 5(g) and (h). The compression of the pulse is not changed significantly as the strength of the CO is varied.
Comparing figures 3 and 5(h), the electron pulse propagating through the focus between C3 and the CO experiences a convergence angle between 2 and 4 mrad, which over the z-axis range of 0.01 m varies between 200 fs and 40 fs. As shown in figure 3, if 2$\sigma$(t) = 100 fs, there is very little temporal influence on the pulse duration at spatial focus. For 2$\sigma$(t) = 10 fs, the duration is perturbed by only a few femtoseconds over this distance.

\subsection{Sample to detector image transport}

As the electron pulse passes the sample plane, it will be scattered by the static structure and dynamic effects associated with the laser pump pulse.. After passing the sample, the duration of the electron pulse has no influence on imaging, rather the pump-probe delay defines time. We verify the imaging capabilities of the instrument, checking that the pulse is not distorted upon propagation from the sample to the detector plane and calculating typical image magnifications.

\begin{figure}
	\centering
	\includegraphics[scale=0.45]{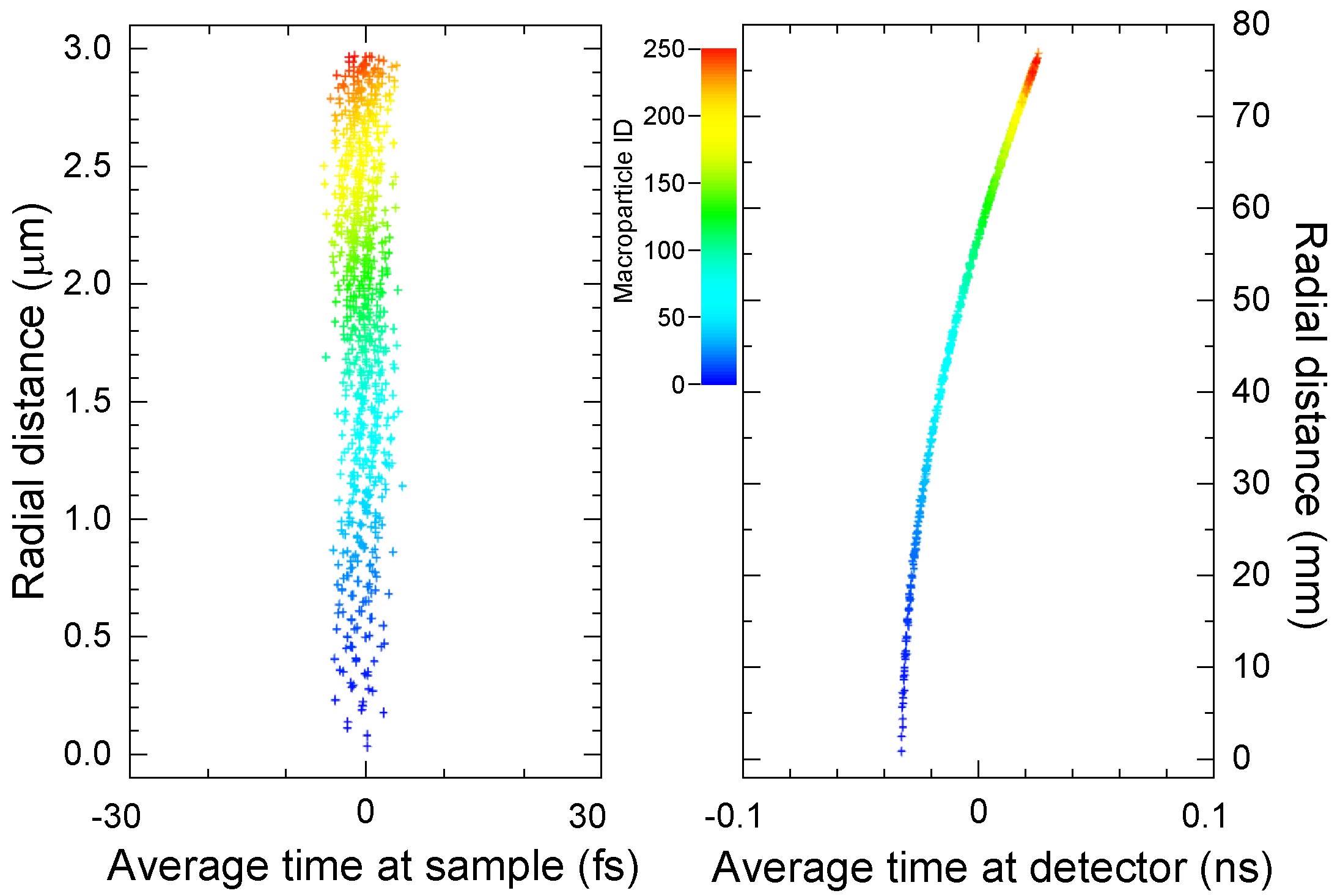}
	\caption{Electron pulse spatial distributions at the (a) sample and (b) detector. In both (a) and (b), the colour scale is macroparticle identification number. (a) Temporal compression indicated by a narrow time of arrival at the sample, flat wavefront and radially increasing particle identification number at the sample. (b) Decompression shown, however the radial distribution of $N_p$ is similar to that of (a). A magnification of greater than $10^{5}$ is predicted. }
	\label{fig:fig6}
\end{figure}

When propagating N macroparticles representing the electron pulse, GPT assigns a particle identification number, Np, which varies between 1 and 250 for these calculations. In figure 6, the variation of Np as a function of radial distance and macroparticle arrival times at both the sample and detector planes is shown. 
In figure 6(a), the electron pulse exhibits a uniform compression to less than 10 fs with a flat wavefront at the sample plane. The apparent structure is due to the Hammersley sequence employed to reduce the shot noise to 1/N. Figure 6(b) shows the electron pulse at the detector plane. The wavefront of the electron pulse is no longer flat, and is approximately 60 picoseconds in duration. However, neither of these factors has a negative impact on the imaging capability of the fs-TEM.

Comparing figure 6(a) and (b), there is a fair agreement between radial position and Np when comparing the pulse at the sample and detector. Minor disagreement is found between macroparticles far from the propagation axis, however considering that the majority of the charge lies close to the propagation axis, this distortion affects a small sample of the pulse population and hence will have minimal impact. Magnification of the order of $10^4$ is also evident upon examination of the length scales of the pulses displayed, fulfilling design criteria (1). 

\subsection{Spatial and dynamic electric field resolution}

To fully exploit the imaging capabilities of the fs-TEM, the detection of electromagnetic fields requires quantification. We employ a 3D electric field to investigate the displacement of electron pulses detectable within the fs-TEM. A custom GPT element deflects the passing electron pulse with an electric field, and removes macroparticles from the pulse if they pass the sample plane within an annular boundary, shown in figure 7(a). The direction and strength of the electric field in the xy plane (mutually perpendicular to the z-axis along which the electron pulse propagates) is indicated by the arrows. The shaded region is the annulus, opaque to electrons, so could be considered thick as compared to the mean free path of the electrons.

\begin{figure*}
	\includegraphics[width=\textwidth]{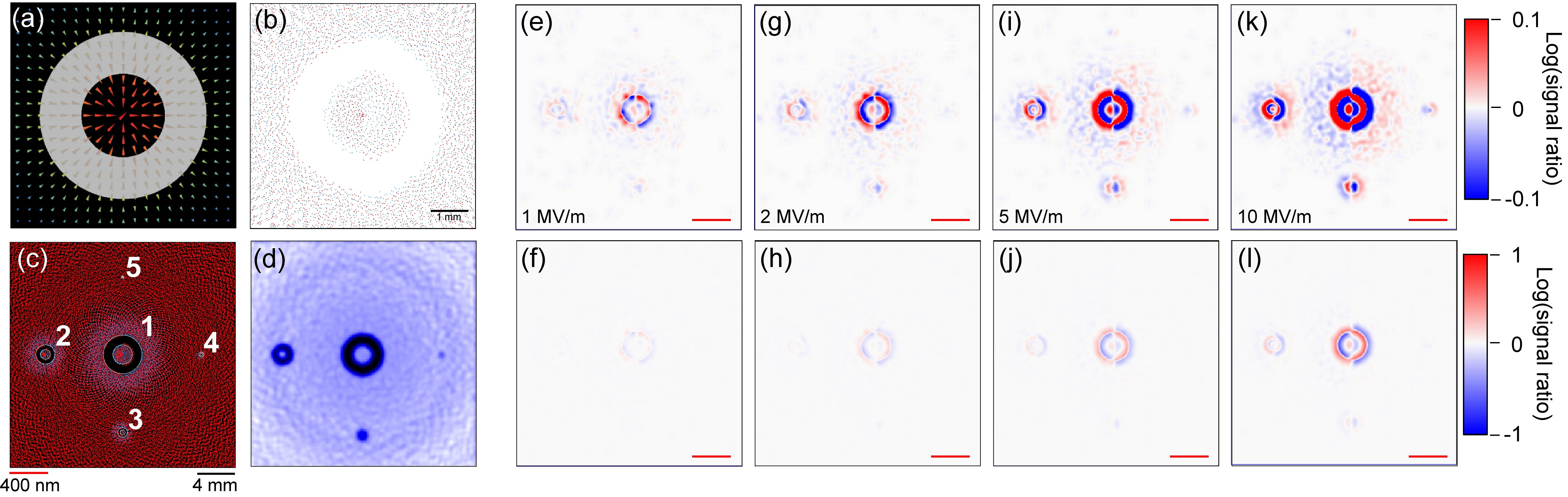}
	\caption{fs-TEM imaging of electric fields at the sample plane. (a) Schematic of the scattering element, with the vector field $E_{scat}$ strength represented by arrow colour and size. (b) Typical macroparticle scattering maps at the detector plane. The green points are the unscattered macroparticles (i.e. $E_{scat}$ = 0) and red are scattered by the element in (a) with $E_{scat}$ = 10 MV/m. (c) Location of five scattering elements of varying size: ($i$ = 1) outer:inner radii of 200:100 nm, ($i$ = 2) 100:50 nm, ($i$ = 3) 50:25 nm, ($i$ = 4) 20:10 nm and ($i$ = 5) 10:5 nm outer and inner radii respectively. (d) Weierstrass transform applied to a macroparticle distribution with a standard deviation $\sigma$ = 350 $\mu$m at the sample plane, recreating the effective fs-TEM image. (e, f) Logarithm of the ratio of the unscattered-scattered fs-TEM images with $E_{scat}$ = 1 MV/m. (g, h) as (e, f) but for  $E_{scat}$ = 2 MV/m. (i, j) $E_{scat}$ = 5 MV/m, and (k, l) $E_{scat}$ = 10 MV/m. Red areas indicate an increase in signal whereas blue areas indicate a deficit, and the red scale bar is 400 nm.}
	\label{fig:fig7}
\end{figure*}

The influence of both the opaque region and localized electric field is shown in figure 7(b), where the point of arrival of the macroparticles at the detector plane are indicated by a point. The green points are the field free case, and the red points indicate the modified trajectories caused by a scattering field $E_{scat}$ = 10 MV/m.

Following the notation convention in figure 7, the geometry of the annuluses from i=1 to i=5 in the format outer radius (nm):inner radius (nm) shown in figure 7(c) are (i = 1) 200:100, (i = 2) 100:50, (i = 3) 50:25, (i = 4) 20:10 and (i = 5) 10:5. The macroparticles representing the electron pulse are propagated though the fs-TEM where they encounter the geometry shown in figure 7(c) and fields shown in figure 7(a). We compare electron pulses that have been deflected ($E_{scat} \neq 0$ ) with those which have not ($E_{scat} = 0$ ) yielding the discrete displacements maps shown in figure 7(b). 

To replicate the performance of the MCP and PS electron detector, a Weierstrass transform (a convolution with a Gaussian kernel) of standard deviation of $\sigma$ = 350 $\mu$m is applied to the discrete macroparticle distributions from electron detection events. This is shown in the conversion from figure 7(c) to (d), and represents a rather blurred and undersampled image. The structures present in the detection event distributions in figures 7(c) and (d) are artefacts from the Hammersley sequence. The Weierstrass transform is constrained by the number of macroparticles used in the numerical propagation, which in the case of figure 7, is N = 20,000. The standard deviation of the Weierstrass transform is an overestimate as compared to what can be achieved with an optimised MCP and PS, which can approach a FWHM of 100 $\mu$m. This artificial spatial blur is necessary as N = 20,000 is a factor of approximately 50 lower than would be wanted for image formation, a limitation applied as the execution time of the space charge calculations scales as N2. 

To create the signal ratio maps shown in figure 7(e) to (l), the $E_{scat} \neq 0$  maps are divided by the $E_{scat} =$ 0 map, and the logarithm taken. Displacement maps for field strengths of $E_{scat} =$ 1, 2, 5 \& 10 MV/m (e \& f, g \& h, i \& j and k \& l respectively) are then displayed on two ranges, here a tenth of an order of magnitude and an order. It is apparent from figure 7 that dynamic electric fields of 1 MV/m will be resolvable. Clearly this depends on the signal to noise ratio of the observation; however in principle such thin and fast acting fields appear observable.

\begin{figure}
	\centering 
	\includegraphics[scale=0.35]{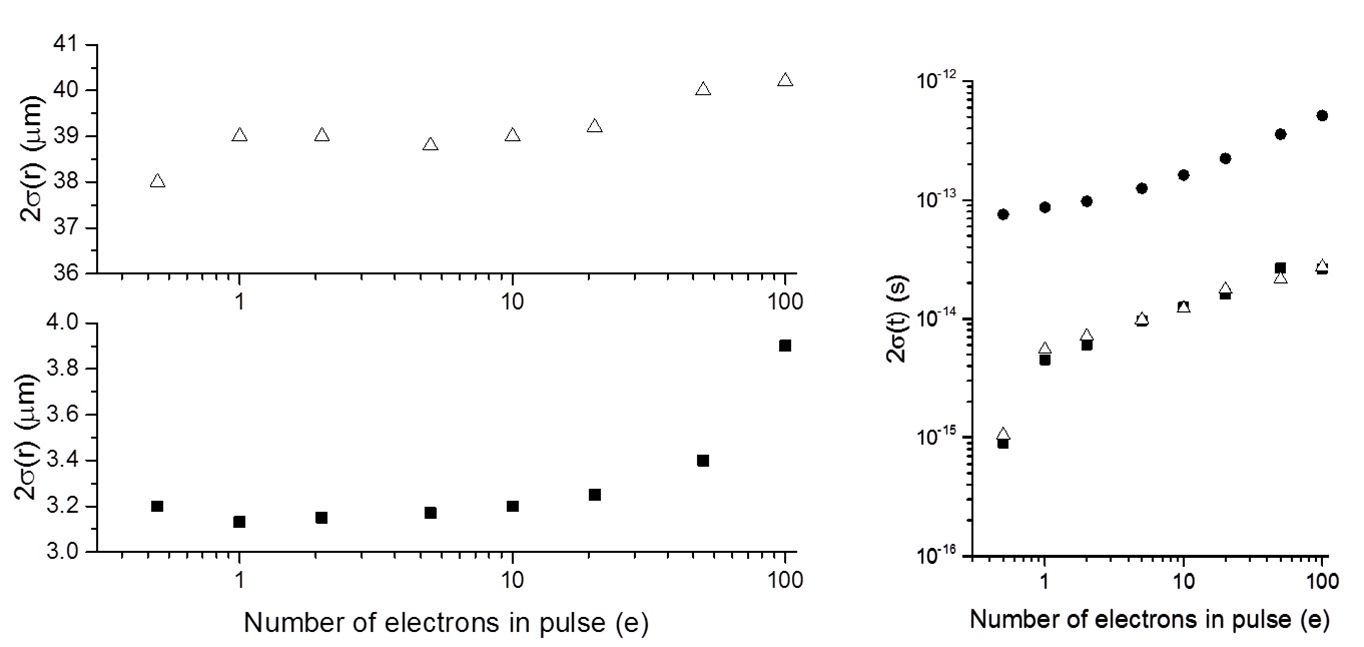}
	\caption{fs-TEM performance as a function of pulse charge under two collimated imaging conditions with anti-parallel C3 and CO excitation. (a) beam size large CO setting. (b) beam size small CO setting. (c) Duration at sample plane for both large and small beams, compared to the pulse duration approaching the resonator.  }
	\label{fig:fig11}
\end{figure}

Figure 8 summarises the performance of the fs-TEM for two illumination modes, corresponding to the largest and smallest 2$\sigma$(r) in figure 5(b), as the number of electrons per pulse is varied. For the operational pulse durations (approximately 10 fs for low tens of electrons per pulse), a small variation in pulse 2$\sigma$(r) $\approx$ 3.2 $\mu$m and $\approx$ 39 $\mu$m. The pulse durations for both radii are displayed upon entrance to the resonator and at the sample plane, indicating the level of compression achieved within the fs-TEM. We find that pulse compression by a factor greater than 10 is achievable, yielding a 2$\sigma$(t) duration of approximately 10 fs, for pulses containing up to 20 electrons. The independent temporal and spatial resolution are seen from the overlap between the larger electron beam (white triangles) and smaller (solid black squares) in figure 8(c) across all number of electrons per pulse investigated.

\section{Discussion}
We present a fs-TEM instrument capable of delivering single to tens of micron beams of electrons, of sub 10 fs duration to the sample through a novel condenser objective lensing setup. Full source to detector beamline simulations are presented, including demonstration of achievable magnification and capability to image dynamic electromagnetic fields at the sample plane. Optimisation and investigation into each beamline component has allowed for a compact, high performance, table top instrument to be designed. Performance of the instrument for a wide variety of electron pulse characteristics has been investigated, allowing one to tailor the electron pulse to the application in mind whilst maximising photo-electron current upon the sample, ensuring viable imaging times. The system is envisioned to run at MHz repetition rates due to the relatively low charge densities of the pulses utilised. All-optical triggering of both the photo-emission process and the excitation of the resonator guarantees electron pulse duration limited temporal resolution.

The length, time and field resolution of the fs-TEM are appropriate to visualise ultrafast processes in the coupled fields of nanophotonics and optical computing, resolve coherent optical and acoustic phonon dynamics in organic and inorganic samples, laser induced magnetism dynamics in nanostructures and thin films and potentially the tracking of charge carrier dynamics in photovoltaics and organic optoelectronic devices.

\section{Methods}
\subsection{Numerical Methods}
To model the electrostatic landscape around the NSMT we utilise the Poisson Superfish package \cite{LANL}.The length scales involved in our geometry, from nanometres at the NSMT apex and significant fractions of a metre between elements necessitate a variable mesh resolution when the geometry is discretized. To avoid discontinuities in the field maps the maximum change in mesh density between two adjacent mesh regions was limited to a factor of two. The numerical accuracy of Poisson was set to $10^{-10}$. 

The General Particle Tracer (GPT) \cite{GPT} package is used for dynamic electron pulse simulations. Utilising an embedded fifth-order Runge-Kutta algorithm with a variable stepsize, GPT numerically calculates electron pulse trajectories through the acceleration region described by the Superfish Poisson field maps and any subsequent beamline components. GPT employs macroparticles to represent an electron pulse, whereby the number of macroparticles and total pulse charge can be varied independently. 250 macroparticles were used for all simulations unless stated otherwise, therefore the statistics of the results presented are independent from the total bunch charge. Convergence testing of the field map resolution is carried out by examining the trajectories and velocities of electron pulses accelerated over the Poisson Superfish field maps as a function of mesh resolution. GPT convergence tests on both the accuracy parameter and the number of macroparticles were conducted, yielding the least computationally expensive means of producing accurate results. Space charge effects are computed using a full 3D point-to-point method calculated from relativistic particle-particle interactions. The accuracy for the numerical propagation is set to $10^{-8}$.

The average energy of electron release from the NSMT is 0.5 eV, with a bandwidth of 0.5 eV, in agreement with previous results \cite{Yanasigawa2011,Kirchner2013}. The influence of bandwidth is indicated by the range bars in figure 1(f) at small z when the electron pulse is in the vicinity of the NSMT.

\subsection{Design Criteria}
(1) The spatial resolution of the instrument should be sufficient to resolve nanostructures, hence an appropriate resolution should be of the order of single to tens of nanometres. Assuming a combined MCP and phosphor screen spatial resolution of 100 $\mu$m, a resolvable feature of 1 nm requires $\times$100k, and 10 nm requires $\times$10k. Taking a typical microchannel plate (MCP) detector radius as 40 mm, $\times$10k magnification would result in a spot size at the sample of 4 $\mu$m. 

(2) To ensure the fs-TEM is capable of resolving the electronic processes discussed, we require a temporal resolution of below 10 fs. This will take the performance of the proposed instrument beyond what is currently possible, with typical optimal temporal resolutions of 100 fs. Utilising a single drive laser system to trigger photoemission, temporal compression and the optical pump of the sample should ensure a jitter down to single femtoseconds \citep{Kealhofer2016}, hence aiming for 10 fs performance is reasonable. 

(3) Providing conditions (1) and (2) can be maintained, delivering the maximum number of electrons per pulse is advantageous but requires care to be taken. Considering a 70 keV TEM with a 10 nA beam that has been spatially focussed to a spot of radius 1 $\mu$m, there will be of the order of $10^{15}$ electrons per m$^3$. Upon comparison with a 10 fs electron pulse containing 10 electrons, electron densities reach $10^{19}$ electrons per m$^3$, and we see that ultrafast multi-electron pulses at spatial foci can reach densities at least $10^4$ times higher than a traditional TEM under comparable conditions. 

(4) All of the electrons emitted from the NSMT should contribute to image formation to ensure reasonable image acquisition times. We acknowledge that employing apertures in the fs-TEM would improve both the spatial (smaller convergence angles) and temporal (minimizing geometric effects) resolution. But, as figure 1(h) shows, generating an electron pulse of hundreds of electrons then propagating through the source along causes a stretch to hundreds of femtoseconds. The temporal stretch experienced in the first few millimetres in a highly apertured pulse would make the requirement of the compressor far more significant, resulting in shorter temporal focal lengths hence compromising magnetic lens and sample positioning. 

(5) Typical drive laser sources are tabletop in scale, and require laboratory facilities such as environmental control, stable AC power, vibration isolation using optical tables, and should operate without generating electric and magnetic noise or transients. We aim for the fs-TEM to be realisable in a tabletop environment, and that cryogenics are not necessary for superconducting magnets.

\subsection{Magnetic Lenses}
All lenses utilised within this study are constructed using single loop solenoids, which act as thin lenses. This allows GPT to treat the resultant magnetic field analytically rather than using a Poisson solver. This is advantageous as single loop solenoids are not computationally expensive and require no approximations. To ensure the single loops employed are a fair analogue, we require that the field strengths generated do not exceed those that might be generated in a typical magnetic lens carrying thousands of amp-turns surrounded by a soft magnetic pole piece with a saturation field of 1.7 T. Saturation of pole pieces in a magnetic lens causes leakage of the field out from the magnetic circuit, degrading the quality of spatial focusing characteristics and often resulting in astigmatic images. Such distortions are expected to have a deleterious effect on the spatio-temporal distribution of charge in of the electron pulse.

An additional consideration is the electrical power dissipated in a magnetic lens. The maximum current applied to the lens depends on the availability of cooling, which we assume will be provided by water circuits. Taking a water flow rate of 2 l/min, an achievable power dissipation of 5 kW m$^{-2}$ K$^{-1}$ is sufficient to allow current densities of up to 50 A mm$^{-2}$ in round wire and up to 200 A mm$^{-2}$ in tape windings. Such current densities allow magnetic fields of a few Tesla to be produced in lenses of a thickness below ten centimetres, which is compatible with the configuration shown in figure 2.

The condenser-objective lens currents for the calculations shown in figure 5 are given in figure 9 for the parallel and antiparallel excitations.

\begin{figure}
	\includegraphics[scale=0.35]{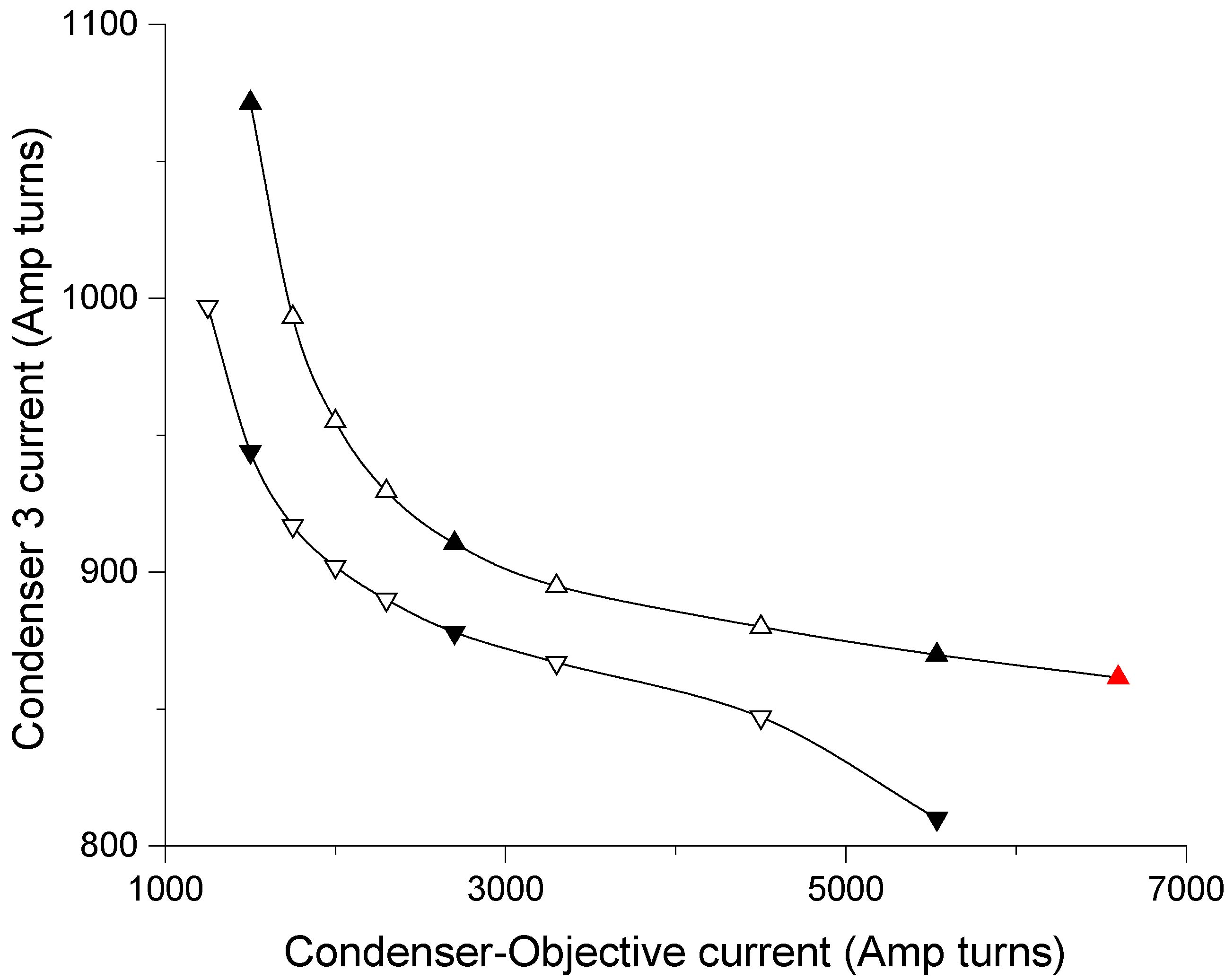}
	\caption{(Condenser-objective and condenser lens 3 currents for the parallel (up triangles) and antiparallel (down triangles) excitation polarities for the results presented in figure 7. The black filled points identify the thick lines in figure 7, and the red filled triangle is the smallest electron beam diameter achieved.}
	\label{fig:fig8}
\end{figure}

\subsection{Temporal Compression}
Our temporal compressor is a Gaussian distribution along the z-axis (electron propagation axis), which represents the electric field produced by a 100 $\mu$m thick aluminium resonator illuminated with 0.3 THz radiation. The Gaussian standard deviation reproduces the field inside the resonator along with a good approximation to the evanescent fields. A sinusoidal temporal variation is applied with a frequency of 0.3 THz, and the phase defined so as to optimize compression. The resonator design only transfers momentum along the z-axis. The electron beam is spatially focussed at the centre of the resonator to a diameter below a micron by the second condenser lens, keeping spatial variation of the compression field across the electron pulse to a minimum, here less than 0.3\%. While our generic compressor has a uniform field in the $xy$ plane, in practice some variation is always possible. 

\subsection{Scattering Field}
The scattering electric field decreases radially in the $xy$ plane from a central maximum of $E_{scat}$ following a Gaussian distribution, $\sigma_r$ which scales with annulus outer radius. The field direction rotates linearly with $\theta$ until a direction reversal at $\theta$ = 0 and $\pi$, with $\theta$ being angle in the $xy$ plane. The scattering field is gaussian along the $z$-axis with a standard deviation $\sigma_z$ = 100 nm. This thickness is chosen so as to be a realistic representation of the distance over which a nanoscale object would exhibit a dynamic field. There is no time dependence to $E_{scat}$, however the convolution of the temporal duration of the electron pulse of 8.2 fs with the 100 nm thickness at 70 keV implies an interaction time with a 2$\sigma$(t) = 9.3 fs.

\subsection{Timing Stabilities}
To what extent the durations predicted in figure 4 can be achieved depends on the stability of the laser pulse energy and pointing, the DC high voltage accelerating the electron pulse, magnetic lens current supplies and mechanical instabilities brought about by vibration and thermal effects throughout the optical and electron transports. RMS laser pulse energies of 0.5$\%$ over days are reported by manufacturers. Micro-radian pointing or sub-micron positional stabilities are common with active stabilization, which coupled to laser optics with focal lengths around 10 cm correspond to intensity fluctuations around 0.3$\%$. Voltage and current supplies have been developed to satisfy stringent requirements in TEM, and are available with part per million stabilities. The most significant noise source appears to be pulse energy, which will influence number of electrons per pulse, and if optical parametric or frequency doubling are used, could also change the duration of the IR drive laser on the NSMT. 

\bibliographystyle{ieeetr}
\bibliography{fstembib}

\end{document}